\pdfoutput=1
\documentclass[11pt]{article}
\usepackage{ieee}
\usepackage{authblk}
\usepackage{epsfig}
\usepackage{comment}
\usepackage{cite}
\usepackage{graphicx}
\usepackage[cmex10]{amsmath}
\usepackage{algorithm}
\usepackage{algorithmic}
\usepackage[tight,footnotesize]{subfigure}
\usepackage{url}
\usepackage{wrapfig}

\begin{document}

\begin{center}

Department of Computer Science and Engineering \\
University of Texas at Arlington \\
Arlington, TX 76019 \\

\begin{figure}[h]
\vspace{-1.5in}
\begin{center}
\includegraphics[width=3.5in, height=4.5in]{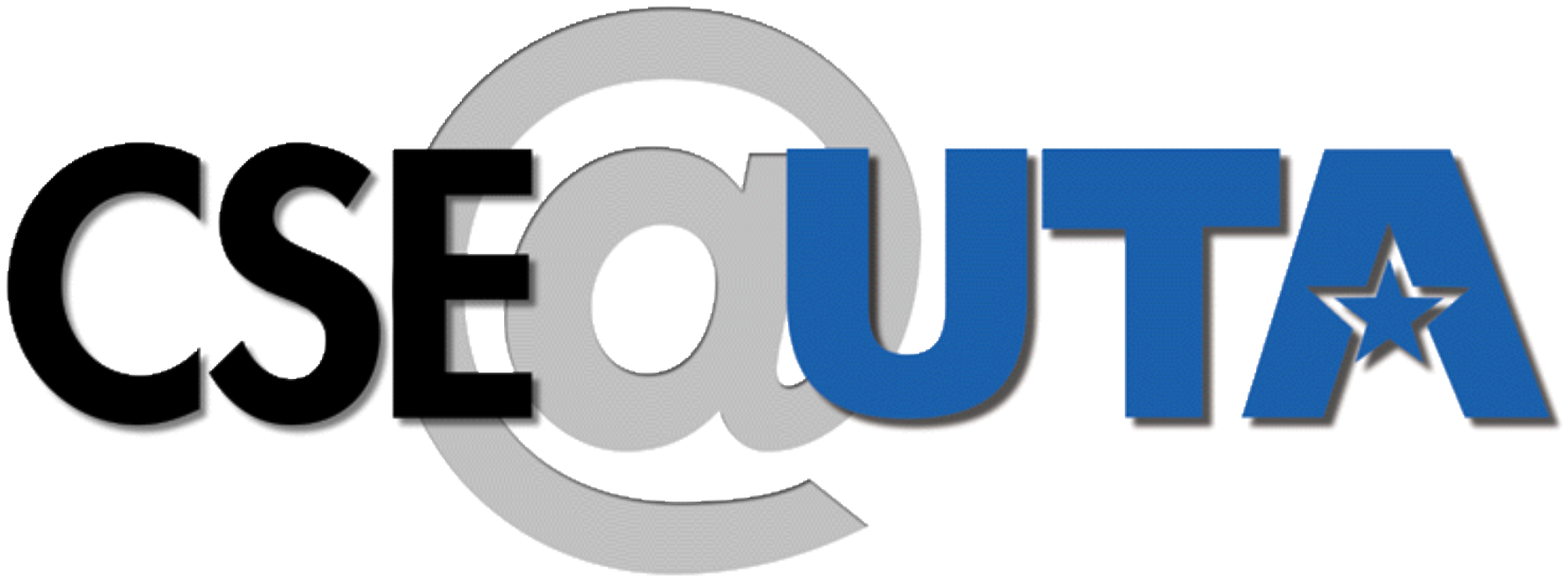}
\end{center}
\end{figure}

\LARGE Scalable Holistic Analysis of Multi-Source Data-Intensive Problems Using Multilayered Networks \\
\vspace{1.5in}
\large Abhishek Santra, Sanjukta Bhowmick and Sharma Chakravarthy \\
\vspace{1.5in}
\large Technical Report \\

\end{center}

\title{Scalable Holistic Analysis of Multi-Source, Data-Intensive Problems Using Multilayered Networks}
\author[1]{Abhishek Santra\thanks{abhishek.santra@mavs.uta.edu}}
\author[2]{Sanjukta Bhowmick\thanks{sbhowmick@unomaha.edu}}
\author[1]{Sharma Chakravarthy\thanks{sharma@cse.uta.edu}}
\affil[1]{IT Lab, CSE Department, University of Texas at Arlington, Texas, USA}
\affil[2]{Department of Computer Science, University of Nebraska at Omaha, Nebraska, USA}

\maketitle

\begin{abstract}

Holistic analysis of many real-world problems are based on data collected from multiple sources contributing to some aspect of that problem. The word fusion has also been used in the literature for such problems involving disparate data types. Holistically understanding traffic patterns, causes of accidents, bombings, terrorist planning and many natural phenomenon such as storms, earthquakes fall into this category. Some may have real-time requirements and some may need to be analyzed after the fact (post-mortem or forensic analysis.) What is common for all these problems is that the amount and types of data associated with the event. Data may also be incomplete and trustworthiness of sources may also vary. Currently, manual and ad-hoc approaches are used in aggregating data in different ways for analyzing and understanding these problems.

In this paper, we approach this problem in a novel way using multilayered networks. We identify features of a central event and propose a network layer for each feature. This approach allows us to study the effect of each feature independently and its impact on the event. We also establish that the proposed approach allows us to compose these features in arbitrary ways (without loss of information) to analyze their combined effect. Additionally, formulation of relationships (e.g., distance measure for a single feature instead of several at the same time) is simpler. Further, computations can be done once on each layer in this approach and reused for mixing and matching the features for aggregate impacts and "what if" scenarios to understand the problem holistically. This has been demonstrated by recreating the communities for the AND-Composed network by using the communities of the individual layers.

Specifically, we propose a representation of disparate data as multilayered network, that can capture the inter-relation between the events and makes it easier to add new information to individual layers as they become available. Further, algorithms have been given for combining multiple layers in any arbitrary manner in order to facilitate the study of the combined effect of different sources. Finally, we present an elegant and low-cost method to combine analytical results from multiple layers, without recomputing the combined layers.

We therefore believe that techniques proposed here make an important contribution to the nascent yet fast growing area of data fusion.

\end{abstract}

\section{Introduction}
\label{sec:introduction}

A critical aspect of big data analysis is identifying how different features, collectively or individually contribute to a central event. From natural phenomena such as storms, earthquakes, to traffic accidents, to premeditated crimes such as terrorist attacks, all events are multifaceted in nature. Multiple data sources capture different perspectives for each event. The central question in analyzing such multifaceted data is to study that how the individual and combinations of sources effect the events for a holistic understanding of the central event.

{\bf Motivation:} As an example, consider the case of traffic accidents. Each traffic accident is tabulated with associated information (also termed features) such as the geographical location, the date, the time when it happened, the light and weather conditions at the time of occurrence, the number of casualties, the number, type and speed of vehicles, type of the locality (urban or rural), type of the road (one way,  roundabout etc.) and the people involved in the accident. The associated data for these features is captured by various sources. Each feature or combination of features tells a different story about the same set of accidents. For example, two accidents even if they occurred during the same time of the day with similar light and weather conditions involving same type and number of vehicles, may lead to different number of casualties due to marked difference in speed at the roundabout.

Therefore, given such a database of traffic accidents, and associated features, if we can identify accidents that occurred primarily due to poor light, that occurred primarily due to bad weather, or those that occurred due to combination of both, then we can determine with respect to per accident location whether to have infrastructure to improve the lighting or to have warning signs due to bad weather or both. Such selective targeting of features, per event, is useful for responding to other problems that involve multiple features. For example, when a disease is treated by a cocktail of different drugs, physicians often manually tune the dosage of different drugs based on the patient\textsc{\char13}s reactions.

{\bf Problem Formulation and Challenges:} Given a dataset for a central event with multiple instances and associated features, the specific problem we want to address in this paper is to group these instances based on the features. To do so, we have to consider all possible subsets of the features, as every subset will bring out a distinct aspect of the central event. For each subset combination and event instance pair, we cluster event instances that occur at the same value of the feature combination. This is clearly a very computationally intensive task, because for $n$ features, we will have $2^n$ possible subsets, leading to $2^n$ clustering problems. Furthermore, these clusterings will have to be recomputed each time new entries are added to the dataset.

{\it In this paper, we present an elegant graph-theoretic technique using which we can efficiently cluster event instances based on different subsets of features.}

{\noindent\bf Outline of Our Approach:} Our approach to this problem is to represent the information as a multilayered network. Network analysis has become a very popular tool for analyzing systems of interrelated entities. The entities (here the event instances) are represented as vertices. Two vertices (event instances) are connected by an edge, currently unweighted and undirected, if the corresponding feature value between them is similar\footnote{It can be argued that weighted edges might provide a more faithful representation; however, here our goal is to simply connect two events that satisfy a certain level of similarity.}. Therefore instances that have similar features will be tightly connected together and form communities. In this paper, we will focus solely on community detection.

Because we have different features, we can create a separate network, each corresponding to a different feature. Such a set of networks, where the vertices are the same, but the connections between them vary, are collectively called {\em multilayered networks}.

Representation of a multi source dataset as a multilayered network, provides several benefits. {\em First}, networks provide an elegant way of representing similar event instances on a per feature basis. Note that although the feature type might vary, from numeric, to nominal, to time, in each network they are canonically defined by edges. {\em Second}, it is relatively easy to combine the features. In most other scenarios, it is difficult to combine features having different types and domains of values. However, in the proposed network-based model the combination can be achieved by simply taking a union or intersection of the edges as needed. Furthermore, as we will show in Section \ref{sec:otherTasks}, the analytical results obtained from the individual networks can be integrated using Boolean operations to obtain the same results that we would have obtained from the combined network. Thus we only need to solve $n$ analytical problems and use these to obtain the results for the rest of the feature combinations. This part is demonstrated using community detection task. {\em Finally}, this representation as multilayered network facilitates handling of new instances as well as features. Not only can new entries be easily added via simple node, edge and/or layer addition, but the results can be updated quickly by simply combining the new result with the old ones via Boolean operations. Moreover, using link prediction algorithms, missing data can also be inferred.

To summarize, our {\bf main contributions} are as follows:
\begin{itemize}
    \item We propose and discuss the benefits of a multilayered network representation for data-intensive events with a large number of features captured by multiple sources (Section~\ref{sec:LayerIntroduction}).
    \item We define composability rules based on Boolean operations to combine multiple features into AND, OR and NOT-composed networks and thus aid in multiple feature based analysis (Section~\ref{sec:LayerComposition}).
   \item We introduce the concept of {\it self preserving} communities. We show that if the communities in the individual networks are self-preserving, then the communities in AND-composed networks can be recreated by intersecting the communities obtained from individual networks. This showcases that it is possible to infer the combined effect of features without generating the respective composed networks, thus reducing modeling complexity and improving efficiency (Section \ref{sec:otherTasks}).
   \item We augment our analytical results with empirical results on a dataset of traffic accidents. We show that it is important to consider all the different subsets of the features, as each of them affects an event in a unique manner. We also empirically show that by composing the communities we can reduce the computational costs of finding communities in the AND-composed networks. (Section \ref{sec:Experiments}).
\end{itemize}

\section{Related Work}
\label{sec:relatedWork}
In this section we provide an overview of the related work including work in data fusion, multilayered networks, and detecting communities in multilayered networks.

{\bf Data Fusion:} The area of data fusion concerns
 combining data from multiple sources (including video \cite{blasch2014context}) to gain a holistic understanding of the situation. The main challenges in achieving high level information fusion \cite{blasch2010high,blasch2012high} is to link information over different collections so that user queries can be answered based on information extracted from images, videos, and correlated with other sources. Here we propose to use a multilayered network approach to fuse data associated with multiple features, of multiple types and obtained from multiple sources. 

{\bf Multilayered Networks:}  Recently, many analytical tasks have used multilayered network~\cite{MultiLayerSurveyKivelaABGMP13} to handle varying interactions among the same set of entities such as co-authorship network in different conferences \cite{MinSubMulLayer2012}, citation network across different topics, interaction network based on calls/bluetooth scans \cite{ClusSpectral2011} and friendship network across different social media platforms. In each of these cases, the relationship among the entities is of the same type, and well-defined. Examples include  whether people work/interact with each other, cite each other or are friends with each other. In contrast,  we are interested in a class of events that are associated with features of different types, each feature providing a unique perspective. There is yet not much work on how different feature types can be combined to generate a multilayered network for representing various relationships among the same set of nodes.

Further, in order to holistically study an event, we also have to study the impact of the combinations of different layers (or features) in the network. Although,  techniques based on information theory have been proposed for multilayer protein-protein interactions~\cite{LayerAggDomenicoNAL14}, this is only for reducing the number of redundant layers through aggregation, but not as a generalized approach for composing different layers  to represent the corresponding combination of features as proposed here.

{\bf Community Detection:} Community detection involves finding groups of tightly connected vertices in a network. This is a well-studied problem in network analysis, and recent work has also looked into community detection algorithms for multilayered networks \cite{CommSurveyKimL15}. Here we propose a novel approach by which communities obtained from individual layers can be easily combined to obtain communities present in the composed multilayered network. To the best of our knowledge, this technique of inferring the communities of the combined network from layers of individual communities has not been studied before.

\section{Creating Multilayered Networks}
\label{sec:LayerIntroduction}
In this section we describe how we create the multilayered networks from multi-source datasets. The notations introduced in this section to formalize our definitions are summarized in  Table \ref{notations}.

\begin{table}[h!t]
    \renewcommand{\arraystretch}{1.3}
    \caption{List of notations used for defining the concepts.}
    \label{notations}
    \centering
        \begin{tabular}{|c||c|}
            \hline
                $N_{f}$ & Number of event features \\
            \hline
                $N_{I}$ & Number of event instances \\
            \hline
                $I_{i}$ & The $i^{th}$ event instance \\
            \hline
                $f^{k}$ & The $k^{th}$ event feature \\
            \hline
                $t_{k}$ & $f^{k}$ type $\in$ \emph{\{numeric, nominal, date, time, location\}}\\
            \hline
                $f_i^k$ & Value of $I_i$ for $f^k$ \\
            \hline
                $D_{t_k}(f_i^k$,$f_j^k)$ & Distance between $I_{i}$ and $I_{j}$ based on $f^{k}$ \\
            \hline
                $\tau_{f^k}$ & Threshold value for similarity with respect to $f^{k}$\\
            \hline
                $L(V_{k}, E_{k})$/$L_k$ & The $k^{th}$ layer \\
            \hline
                $V_{k}$ & Set of nodes in the $k^{th}$ layer \\
            \hline
                $E_{k}$ & Set of edges in the $k^{th}$ layer \\
            \hline
                $(u_{i}^{k}, u_{j}^{k})$ & An edge between $I_i$ and $I_j$ in the $k^{th}$ layer \\
            \hline
        \end{tabular}
\end{table}

{\bf Multi-Source Datasets:}
Many events are associated with multiple features (or attributes). For example, an accident scene can be described, by several features including light conditions, weather conditions, road conditions, date, etc. Therefore, each event can be described as a tuple of features. Formally, if $N_I$ is the total number of event instances and $N_f$ is the total number of event features, then in general the $i^{th}$ event instance, $I_i$, can be represented as an $N_f$-tuple, shown in equation \ref{eq:instanceRep}, where $f_i^j$ is the $j^{th}$ feature's value.

            \begin{equation}
            \label{eq:instanceRep}
                I_i = <f_i^1, f_i^2, ..., f_i^{N_f}> \forall i \in \{1, N_I\}
            \end{equation}
            
We define {\em distance metric} as the measure of similarity between two event instances. The distance metric is denoted by $D_{t_k}(f_i^k, f_j^k)$ and represents the distance between the $i^{th}$ and the $j^{th}$ instances with respect to the $k^{th}$ event feature, which is of type $t_k$. For each type of feature, multiple distance metrics are possible. Thus, the sample distance measure $D_{t_k}(f_i^k, f_j^k)$ for the different feature types considered for this paper are as follows. Note that we define the distance such that lower distance indicates higher similarity.
\begin{itemize}
    \item \emph{Numeric ($t_k$ {= numeric})}: Features such as number of casualties caused by the accident and the speed limit of the road where the accident occurred, whose values correspond to integers or floating point numbers fall under this category. Equation \ref{eq:numDistance} defines the distance metric as the absolute difference between the values of the features.
            \begin{equation}
            \label{eq:numDistance}
                D_{t_k}(f_i^k, f_j^k) = \lvert f_i^k - f_j^k \rvert
            \end{equation}

    \item \emph{Nominal ($t_k$ {= nominal})}: Nominal features have a fixed discrete set of values. For example, the domain for the feature capturing road surface conditions is \{dry, wet/damp, flood, snow, frost/ice, oil, mud\}. For such features, Equation \ref{eq:nomDistance} states that the distance metric is given as 0 if there is an exact match and undefined (denoted as $\phi$) if they do not match.
            \begin{equation}
            \label{eq:nomDistance}
                D_{t_k}(f_i^k, f_j^k) =
                    \begin{cases}
                        0 & \text{if } f_i^k = f_j^k \\
                        \phi & \text{if } f_i^k \neq f_j^k
                    \end{cases}
            \end{equation}

    \item \emph{Date ($t_k$ {= date})}: Certain features will depict the date of occurrence of the event instance. The distance measure is the number of  days between the occurrences of two instances.
            \begin{equation}
            \label{eq:dateDistance}
                \begin{aligned}
                    D_{t_k}(f_i^k, f_j^k) = daysBetween(f_i^k, f_j^k)
                \end{aligned}
            \end{equation}

    \item \emph{Time ($t_k$ {= time})}: This feature gives the exact time of the occurrence in hours (HH), minutes (MM) and seconds. To compute the distance metric we divide the day into 48 intervals of 30 minutes each from [0000-0030) to [2330-0000). We assume that two events taking place around the same time interval may be similar in nature, even if they happen on two different dates. For example, a set of accidents may be similar because they occur during the evening rush hour on any of the weekdays. Thus, Equation \ref{eq:timeDistance} states that for a time based feature, the number of 30 minute intervals between the occurrences of two given instances, will be used as the distance measure.
            \begin{equation}
            \label{eq:timeDistance}
                \begin{aligned}
                    D_{t_k}(f_i^k, f_j^k) &= \lVert [ 2*f_i^{k_{HH}} + 1 + \lfloor{f_i^{k_{MM}}/30}\rfloor ] \\
                                    &- [ 2*f_j^{k_{HH}} + 1 + \lfloor{f_j^{k_{MM}}/30}\rfloor ] \rVert
                \end{aligned}
            \end{equation}

    \item \emph{Geographical Location ($t_k$ {= location})}: The geographical location of an event's occurrence is given by its latitude value (LAT) and longitude value (LONG). We use the Haversine formula (\cite{HaversineFormulaWiki}) that calculates great-circle distance between any two points on the earth's spherical surface to define the distance metric in Equation \ref{eq:locDistance} for the location based features, considering R to be the radius of the earth.
            \begin{equation}
            \label{eq:locDistance}
                \begin{aligned}
                    D_{t_k}(f_i^k, f_j^k) &= 2R \arcsin * ( \sqrt ( \sin^2(\frac{f_i^{k_{LAT}} - f_j^{k_{LAT}}}{2}) \\
                    &+ \cos(f_i^{k_{LAT}}) * \cos(f_j^{k_{LAT}}) * \sin^2(\frac{f_i^{k_{LONG}} - f_j^{k_{LONG}}}{2})) )
                \end{aligned}
            \end{equation}
            \end{itemize}

In addition to the types listed above, other types such as videos, images, audio files, tweets, SMS etc. can also enhance the description of the event. Currently we are not considering these types for this paper.

{\bf Creating the Multilayered Network:} Based on the distance metric we now represent the dataset as a multilayered network (or graph). For  a given feature, we say that a pair of event instances are similar if their distance metric is below a specified threshold. We create a separate network for each feature. The instances are represented as vertices in the network. Two vertices are connected in a network, if for that corresponding feature, they are similar. Therefore to create a layer of the network based on a specific feature, we need the following information:

\begin{itemize}
    \item A set, \emph{I}, of all the event instances such that $I = \{I_1, I_2, ..., I_{N_I}\}$
    \item The type of the $i^{th}$ feature, $t_i$. For the current paper, we have considered $t_i$ $\in$ $\{$\emph{numeric, nominal, date, time, location}$\}$.
    \item The metric, $D_{t_i}(f_m^i, f_n^i)$, defined to calculate the distance between any two event instances, $I_m$ and $I_n$.
    \item A specified threshold value, $\tau_{f^i}$, that dictates the similarity between any two instances with respect to the $i^{th}$ feature.
\end{itemize}

Formally, in the $i^{th}$ layer, the $j^{th}$ instance, $I_j$, will be depicted by the $j^{th}$ vertex, $u_j^i$. The presence of an undirected and unweighted edge in this layer, ($u_j^i$, $u_k^i$), will depict that the $j^{th}$ and the $k^{th}$ instances are similar to each other with respect to the $i^{th}$ feature. Each network layer can be uniquely defined by the feature it represents, and will be denoted as $L(V_{i}, E_{i})$ or $L_i$. Note that every layer will have the same set of nodes, but different set of edges.

\begin{figure}[h]
    \centering
    \includegraphics[width=0.5\textwidth]{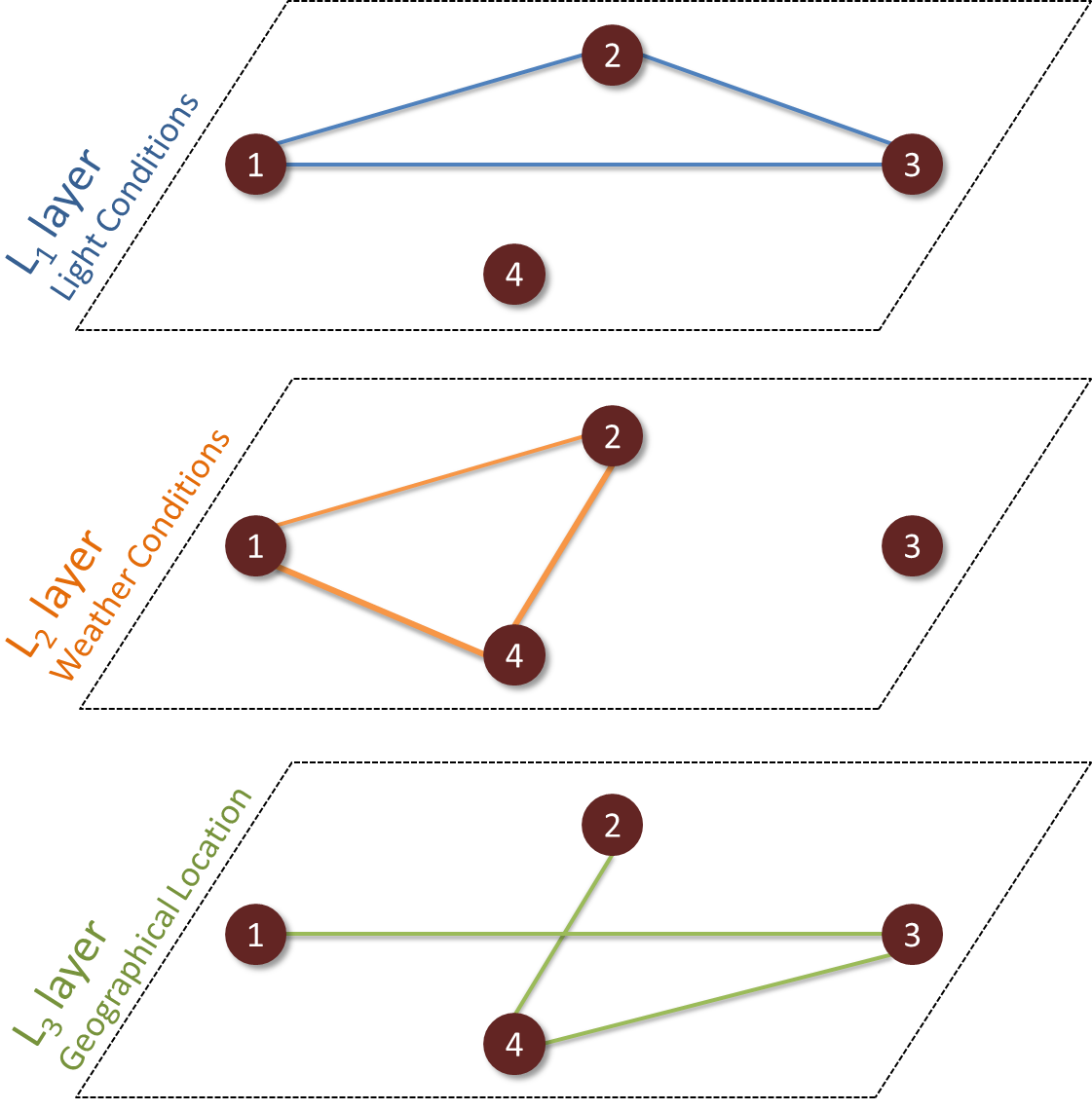}
    \caption{Snapshot of Multilayered network for the accident event}
    \label{fig:multiLayerExample:Accident}
\end{figure}

Figure \ref{fig:multiLayerExample:Accident} shows a multilayered network for four accident instances, denoted by four nodes numbered from 1 to 4. Similarity among the accidents is considered with respect to two nominal features - Light Conditions and Weather Conditions and one location based feature - (Latitude, Longitude), with the threshold value for distance metric being 2 miles. Note that the connectivity of each layer in the network is different, highlighting the unique perspective of each feature. For instance, accident2 and accident3 had the same light conditions when they occurred, but didn't share the weather conditions. This small snapshot shows that every feature tells a different perspective about the relationship among the accidents, thus supporting the relevance of analyzing any event in a perspective-wise manner.

Representing multi-featured datasets using multilayered networks provides the following benefits:

\begin{itemize}
    \item \emph{Ease of handling the dataset incrementally.} The multilayered network representation makes it easy to add or delete new entries and features into the data set. This is because each layer is generated independently from other layers and hence only the affected layer has to be changed through the addition or deletion of nodes and/or edges. As we will see in the next section, even combining the features is made more effective due to the multilayered approach.

    \item \emph{Identifying importance of features on the central event.} The multilayered framework allows us to analyze the contribution of individual or combined features. The importance of a feature can be measured by factors such as  the edge density of the network, the number of connected components and the community structure. These measurements can help us order the features in terms of their importance.

    \item \emph{Determining the strength of a relationships.} The network-based representation allows us to easily identify the strength of the relationships between event instances. For example, the instances that have an edge between them across multiple layers are more strongly related than if they have an edge in only one layer.

    \item \emph{Inferring feature dependencies and missing instance-instance relationships.} Every layer has the same set of nodes but different set of edges. Thus, based on the edge connectivity a correlation can be identified among the features. In case of missing feature values, these inferred correlations among the features will aid the link prediction algorithms to infer the missing relationships.

     \item \emph{Efficient computation of the effect of multiple features.}  Combining different subsets of $n$ features requires us to solve $2^n$ separate problems. However, the network-based representation allows us to easily combine different layers in any arbitrary manner using Boolean operations (Section \ref{sec:LayerComposition}). Moreover, results of an analytical task for a combined network can be obtained by only using the results from the individual layers. This aspect is discussed in more detail with respect to the community detection task in Section~\ref{sec:otherTasks}.
\end{itemize}

\section{Layer Composition Through Boolean Operations}
\label{sec:LayerComposition}

Each layer in the network provides information of how a single feature effects the event instances. However, it is extremely pivotal to study the effect of the combination of features, as each combination presents a new perspective. For a set of $n$ features, a total of $2^{n}$ different feature combinations are possible.

For example, for an accident dataset with two features, light and weather, the combinations can be generated in the following 4 ways: i) both light and weather, ii) either light or weather,  iii) light and not weather (only light) \emph{or} iv)  weather and not light (only weather).

However, creating a network with multiple features leads to an additional challenge of how to compute the distance metric for a combination of features. To address this challenge we propose the use of fundamental Boolean algebraic operators. Therefore, with respect to the multilayered network, the distance measure criterion based on k features should correspond to the k-layer combination scheme - $(L_{i1}$ $\theta$ $L_{i2}$ $\theta$ ... $\theta L_{ik})$, where $\theta$ represents the type of boolean operator, i.e. AND (Section \ref{subsec:ANDLayer}), OR (Section \ref{subsec:ORLayer}) and NOT (Section \ref{subsec:NOTLayer}).

\subsection{AND Composition}
\label{subsec:ANDLayer}
The AND composition over a set of layers includes an edge only if it occurs in {\em all} the layers. This indicates that the pair of event instances connected by the edges satisfies the threshold parameter for all the required features.

\begin{figure}[h]
    \centering
    \includegraphics[width=0.5\textwidth]{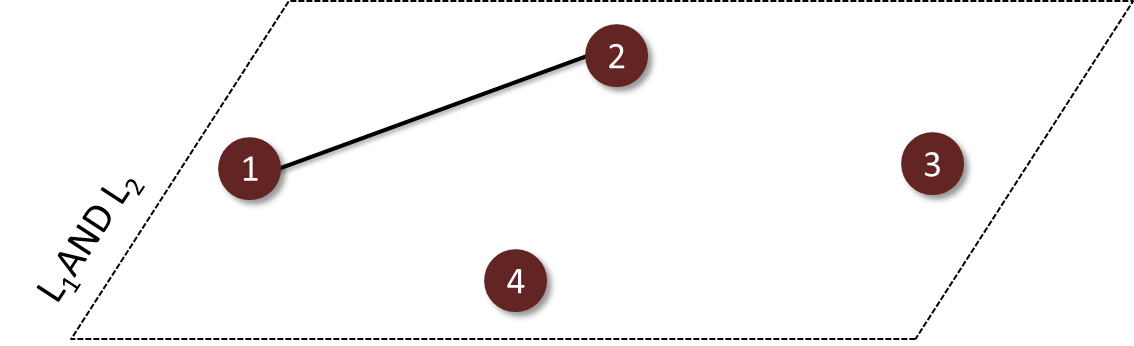}
    \caption{2-layer AND composition applied on the Light layer and Weather layer present in Figure \ref{fig:multiLayerExample:Accident}}
    \label{fig:ANDCompositions:Accident}
\end{figure}

In Figure \ref{fig:multiLayerExample:Accident}, for $L_1$ and $L_2$, the set of edges $E_1$ and $E_2$ depict the accident pairs that are similar based on light and weather, respectively. Figure \ref{fig:ANDCompositions:Accident} depicts the $L_1$ \emph{AND} $L_2$ composition, that contains only edges that are present in both $E_1$ and $E_2$.

Formally, the AND composition of two layers, \emph{L(}$V_{i}$\emph{, }$E_{i})$, \emph{L(}$V_{j}$\emph{, }$E_{j})$, will produce the composed layer \emph{L(}$V_{iANDj}$\emph{, }$E_{iANDj})$. A representative vertex, $u_m^{iANDj}$, is added to the set of vertices, $V_{iANDj}$, for each event instance $I_m$. For any event instance pair, $I_m$ and $I_n$, if an edge exists between their representative vertices in both layer $L_i$ and layer $L_j$, then an edge, $(u_m^{iANDj}, u_n^{iANDj})$ becomes a part of the set of edges, $E_{iANDj}$. The steps for 2-layer AND composition are given in Algorithm \ref{algo:ANDComposer}.

\begin{algorithm}
\caption{Algorithm for AND composition}
\label{algo:ANDComposer}
\begin{algorithmic}[1]
   \REQUIRE $<L(V_{i}, E_{i})$, $L(V_{j}, E_{j})>$, $V_{iANDj}$ = $\emptyset$, $E_{iANDj}$ $= \emptyset$

   \FORALL {$u_m^{i} \in V_{i}$}
       \STATE $V_{iANDj} \gets V_{iANDj} \cup u_m^{iANDj}$
   \ENDFOR

   \FORALL {$ u_m^{iANDj}, u_n^{iANDj} \in V_{iANDj}, m > n$}
       \IF {$(u_m^{i}, u_n^{i}) \in E_{i}$ AND $(u_m^{j}, u_n^{j}) \in E_{j}$}
           \STATE $E_{iANDj} \gets E_{iANDj} \cup (u_m^{iANDj}, u_n^{iANDj})$
       \ENDIF
   \ENDFOR

\end{algorithmic}
 \end{algorithm}

A k-layer AND composed network ($L_{{{AND}_{j = 1}^{k}}(ij)}$) indicates that when combining layers $(L_{i1}$ \emph{AND} $L_{i2})$ \emph{AND} $L_{i3})$ ... \emph{AND} $L_{ik})$, a pair of event instances will have an edge between them if $D_{t_{ij}}(f_m^{ij}, f_n^{ij})$ $\le$ $\tau_{f^{ij}}$, for \emph{every} $j \in [1, k]$. Equation \ref{eq:ANDEdgeSetLimits} shows that the number of edges in an AND composed layer will be bounded by the number of edges in layer with the lowest number of connections, since the composition is formed by an \emph{intersection of edges}.

\begin{equation}
    \label{eq:ANDEdgeSetLimits}
        0 \le \lvert E_{{{AND}_{j = 1}^{k}}(ij)} \rvert \le \min_{\forall j \in [1, k]} \lvert E_{ij} \rvert
\end{equation}

\subsection{OR Composition}
\label{subsec:ORLayer}
The OR composition over a set of layers includes an edge if it occurs in any one of the constituent layers. This indicates that the pair of event instances connected by the edges satisfies the threshold parameters for at least one of the features.

\begin{figure}[h]
    \centering
    \includegraphics[width=0.5\textwidth]{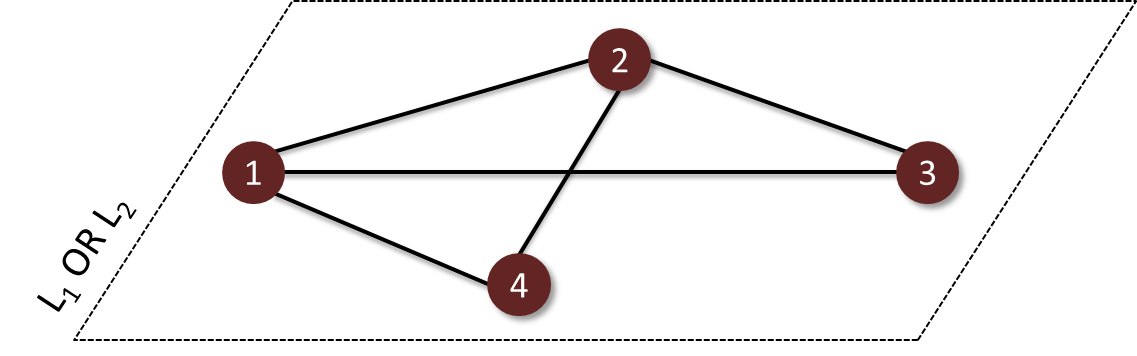}
    \caption{2-layer OR composition applied on the Light layer and Weather layer present in Figure \ref{fig:multiLayerExample:Accident}}
    \label{fig:ORCompositions:Accident}
\end{figure}

For example, for the light and weather based layers in Figure \ref{fig:multiLayerExample:Accident}, the expected result for $L_1$ \emph{OR} $L_2$ composition is a set, which contains edges present in either $E_1$ or $E_2$, or both, as shown in Figure \ref{fig:ORCompositions:Accident}.

The methodology to perform OR composition is similar to the AND Composition and is given by Algorithm \ref{algo:ORComposer}. For the same individual layers considered in Section \ref{subsec:ANDLayer}, the OR composed layer will be $L(V_{iORj}, E_{iORj})$. For every event instance $I_m$, the set $V_{iORj}$ will contain its representative vertex, $u_m^{iORj}$. An edge, $(u_m^{iORj}, u_n^{iORj})$ will be introduced in this composed layer, if the representative vertices of $I_m$ and $I_n$, have an edge between them in either layer $L_i$ or layer $L_j$.

\begin{algorithm}
\caption{Algorithm for OR composition}
\label{algo:ORComposer}
\begin{algorithmic}[1]
   \REQUIRE $<L(V_{i}, E_{i})$, $L(V_{j}, E_{j})>$, $V_{iORj}$ = $\emptyset$, $E_{iORj}$ $= \emptyset$

   \FORALL {$u_m^{i} \in V_{i}$}
       \STATE $V_{iORj} \gets V_{iORj} \cup u_m^{iORj}$
   \ENDFOR

   \FORALL {$ u_m^{iORj}, u_n^{iORj} \in V_{iORj}, m > n$}
       \IF {$(u_m^{i}, u_n^{i}) \in E_{i}$ OR $(u_m^{j}, u_n^{j}) \in E_{j}$}
           \STATE $E_{iORj} \gets E_{iORj} \cup (u_m^{iORj}, u_n^{iORj})$
       \ENDIF
   \ENDFOR

\end{algorithmic}
\end{algorithm}

A k-layer OR composed network ($L_{{{OR}_{j = 1}^{k}}(ij)}$) indicates that when combining layers $(L_{i1}$ \emph{OR} $L_{i2})$ \emph{OR} $L_{i3})$ ... \emph{OR} $L_{ik})$, a pair of event instances will have an edge between them if $D_{t_{ij}}(f_m^{ij}, f_n^{ij})$ $\le$ $\tau_{f^{ij}}$, for \emph{at least one} $j \in [1, k]$. Since the composition is formed by an \emph{union of edges}, the number of edges in an OR composed layer will be bounded by the total number of edges in all the constituent layers, which is shown in Equation \ref{eq:OREdgeSetLimits}. 

\begin{equation}
    \label{eq:OREdgeSetLimits}
        \max_{\forall j \in [1, k]} \lvert E_{ij} \rvert \le \lvert E_{{{OR}_{j = 1}^{k}}(ij)} \rvert \le \frac{N_f(N_f - 1)}{2}
\end{equation}

\subsection{NOT Composition}
\label{subsec:NOTLayer}
\begin{figure}
    \centering
    \includegraphics[width=0.5\textwidth]{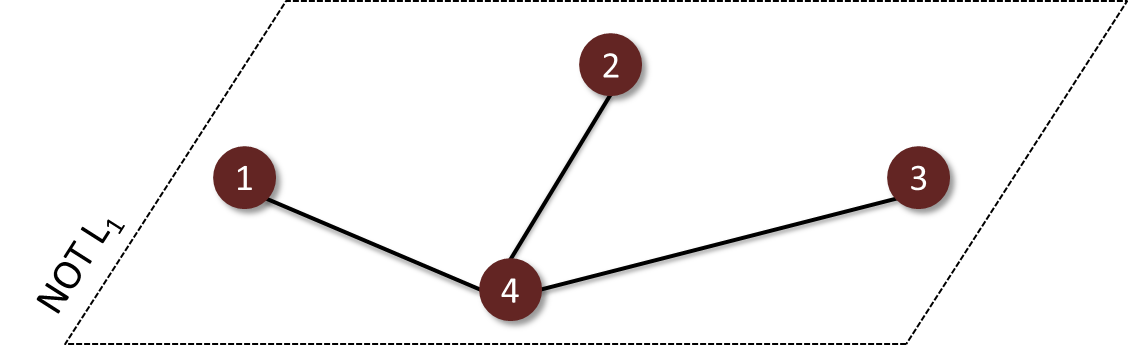}
    \caption{NOT composition of the Light layer present in Figure \ref{fig:multiLayerExample:Accident}}
    \label{fig:NOTCompositions:Accident}
\end{figure}

The NOT composition models the complement of a feature. Thus, NOT composition for the $k^{th}$ layer will generate a network where the existence of an edge will imply that the accident pair does not satisfy the threshold parameter for the $k^{th}$ feature.  For example, for the light based layer in Figure \ref{fig:multiLayerExample:Accident}, the expected result for \emph{NOT} $L_1$ composition is shown in Figure \ref{fig:NOTCompositions:Accident}.

It can be observed in Algorithm \ref{algo:NOTComposer} that unlike the AND and OR composition, the NOT composition is applied on a single layer. The NOT of the $k^{th}$ layer will be a new layer $L(V_{k'}, E_{k'})$, where $V_{k'}$ contains a representative vertex, $u_m^{k'}$, for each event instance, $I_m$. For any two event instances, $I_m$ and $I_n$, an edge $(u_m^{k'}, u_n^{k'})$ is introduced if the representative nodes of these instances do not contain an edge between them in the original layer.

\begin{algorithm}
\caption{Algorithm for NOT composition}
\label{algo:NOTComposer}
\begin{algorithmic}[1]
   \REQUIRE $L(V_{k}, E_{k})$, $V_{k'}$ = $\emptyset$, $E_{k'}$ = $\emptyset$

   \FORALL {$u_m^{k} \in V_{k}$}
       \STATE $V_{k'} \gets V_{k'} \cup u_m^{k'}$
   \ENDFOR

   \FORALL {$ u_m^{k'}, u_n^{k'} \in V_{k'}, m > n$}
       \IF {$(u_m^{k}, u_n^{k}) \notin E_{k}$}
           \STATE $E_{k'} \gets E_{k'} \cup (u_m^{k'}, u_n^{k'})$
       \ENDIF
   \ENDFOR

\end{algorithmic}
\end{algorithm}

The set of edges for this layer, $E_{k'}$, will correspond to the \emph{complement of the set of edges in the $k^{th} layer$}. Therefore, for any two instances, the existence of an edge in $L_{k'}$ depicts that the condition $D_{t_{k}}(f_m^{k}, f_n^{k}) > \tau_{f^{k}}$ is satisfied. From Equation \ref{eq:NOTEdgeSetLimits} it can be concluded that the density of the NOT composed layer will be inversely proportional to the density of the original layer.

\begin{equation}
    \label{eq:NOTEdgeSetLimits}
        \lvert E_{k'} \rvert = \frac{N_f(N_f - 1)}{2} - \lvert E_{k} \rvert
\end{equation}

\begin{figure}
    \centering
    \includegraphics[width=0.5\textwidth]{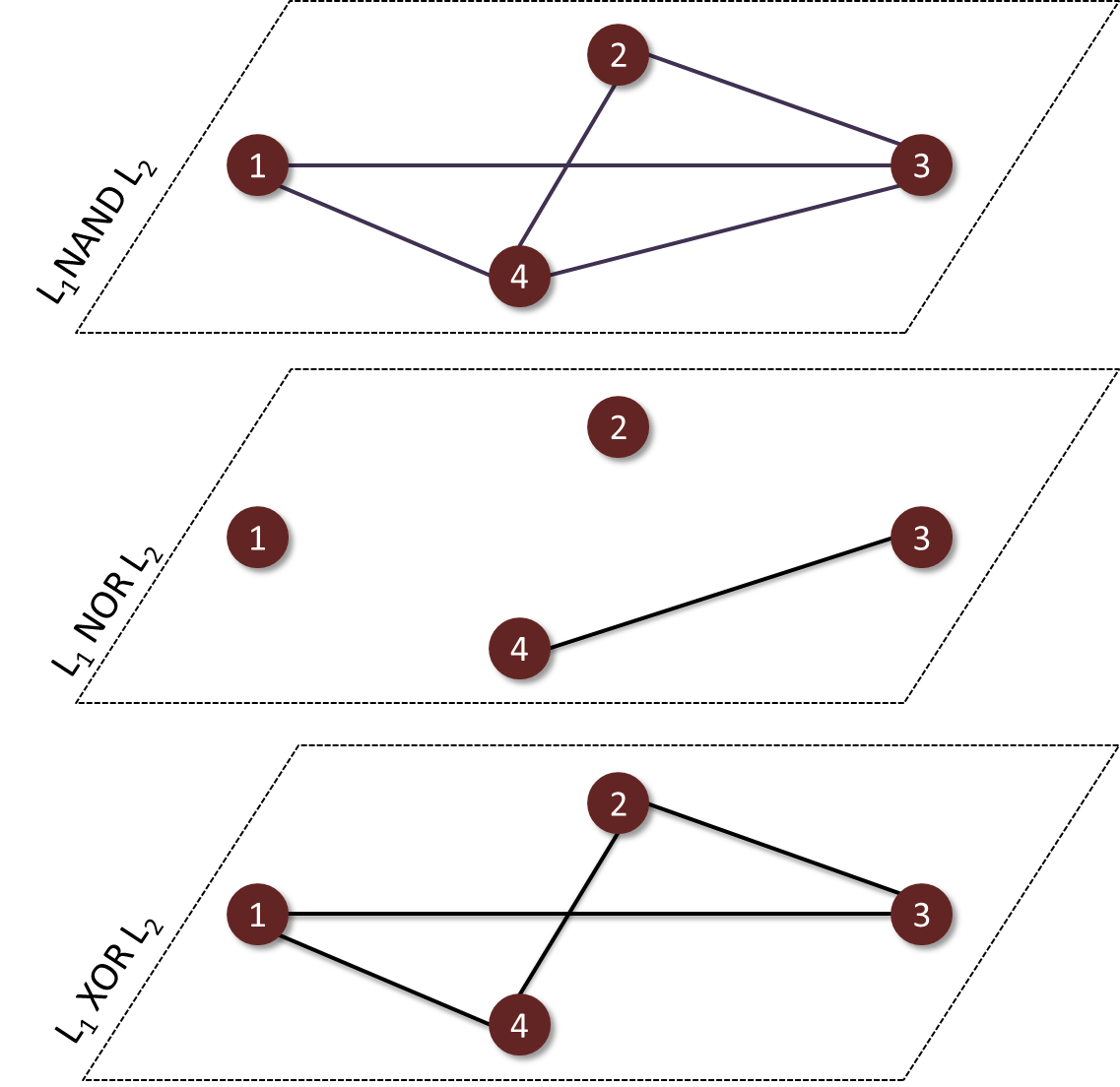}
    \caption{Complex compositions for the layers shown in Figure \ref{fig:multiLayerExample:Accident}}
    \label{fig:otherCompositions}
\end{figure}

{\bf Complex Composition of Layers:} Primitive Boolean operations can be used to create more complex compositions, as shown in Table \ref{table:otherCompositions} and Figure \ref{fig:otherCompositions}. Since these layer compositions are based on Boolean algebra, they will also obey the associative, commutative, distributive and De Morgan's laws, as displayed in Table \ref{table:layerProps}. Using these properties, any complex layer composition of layers can be expressed using the defined  AND, OR, and NOT operations.

\begin{table}[h!t]
    \renewcommand{\arraystretch}{1.3}
    \caption{Complex Layer Compositions}
    \label{table:otherCompositions}
    \centering
        \begin{tabular}{|c||c|}
            \hline
                $L_1$ \emph{NAND} $L_2$ & \emph{NOT} $(L_1$ \emph{AND} $L_2)$ \\
            \hline
                $L_1$ \emph{NOR} $L_2$ & \emph{NOT} $(L_1$ \emph{OR} $L_2)$ \\
            \hline
                $L_1$ \emph{XOR} $L_2$ & $(L_1$ \emph{AND (NOT } $L_2))$ \emph{OR} $(\emph{(NOT } L_1)$ \emph{AND} $L_2)$\\
            \hline
        \end{tabular}
\end{table}

\begin{table}[!t]
    \renewcommand{\arraystretch}{1.3}
    \caption{Layer Composition Properties}
    \label{table:layerProps}
    \centering
        \begin{tabular}{|c||c|}
            \hline
                Commutativity & $L_i$ $\theta_1$ $L_j$ $\equiv$ $L_j$ $\theta_1$ $L_i$ \\
            \hline
                Associativity & $(L_i$ $\theta_1$ $L_j)$ $\theta_1$ $L_k$ $\equiv$ $L_i$ $\theta_1$ $(L_j$ $\theta_1$ $L_k)$ \\
            \hline
                Distributivity & $L_i$ $\theta_1$ $(L_j$ $\theta_2$ $L_k)$ $\equiv$ $(L_i$ $\theta_1$ $L_j)$ $\theta_2$ $(L_i$ $\theta_1$ $L_k)$\\
            \hline
                De Morgan's & \emph{NOT (}$L_i$ \emph{AND} $L_j)$ $\equiv$ \emph{(NOT)}$L_i$ \emph{OR (NOT)}$L_j$ \\
                            & \emph{NOT (}$L_i$ \emph{OR} $L_j)$ $\equiv$ \emph{(NOT)}$L_i$ \emph{AND (NOT)}$L_j$ \\
            \hline
            \hline
                            & where, $L_i$, $L_j$, $L_k$: basic/composed layers \\
                            & $\theta_1$, $\theta_2$ $\in$ \{AND, OR\}\\
            \hline
        \end{tabular}
\end{table}

This section showed how the individual layers in the multi-layered framework can be combined in various ways in order to produce new layers, each presenting an interesting perspective of looking into the relationship among the event instances. In this way, this architecture allows anyone to analyse the impact of multiple features on the central event. 

\section{Combining Analytical Results Using Boolean Functions}
\label{sec:otherTasks}
For a given dataset, one of the primary tasks that we are interested in this paper is to show that analytical results with respect to a combination of features can be inferred by just using the results obtained with individual features. To illustrate this, we chose the analytical task as the clustering of event instances (accidents in our example) based on the single or combined set of features. In the network context, this is equivalent to identifying groups of tightly connected vertices or communities~\cite{CommFortunatoC09,CommNewGir04}.

Although we presented an elegant method for combining individual layers of networks using Boolean operations, we still have to find the communities in these different combined networks. Thus we have to solve $2^n$ separate community detection problems.

In this section, we analytically show that if communities follow certain characteristics then we can reproduce the communities of the composed networks. This reduces the memory requirements since we do not have to load each of the separate composed networks into memory and also reduces computational time because we can recreate the communities using simple Boolean operations, rather than expensive community detection methods.

{\bf Recreating Communities in AND-composed Networks:}
We first introduce the concept of  {\em self preserving} communities. A community is self preserving if the vertices in the community are so strongly connected such that even if only a subset of connected vertices remain in a community, they will form a smaller community rather than joining an existing larger community.

Formally, consider a network $G$, that has a community whose vertices are given by the set $C_v$. Now consider the network induced by a subset of vertices  $C^S_v \in C_v$ and all other vertices that are not in $C_v$. If the vertices in $C^S_v$ form a community by themselves, for any subset $C^S_V$ of $C_v$, where $\|C^S_v\| \ge 3$ and the vertices in $C^S_v$ are connected, then community $C_v$ is self preserving.

Now, consider two networks $G1$ and $G2$ that have the same set of vertices, but different set of edges. Moreover, both networks have only self-preserving communities. Now consider the AND-composition of $G1$ and $G2$, $G_{AND}$. Only edges that are in both $G1$ and $G2$ will be in the AND-composed network. Therefore the communities formed in the AND-composed network will be based on a subset of edges from $G1$ and $G2$. Since both $G1$ and $G2$ have self preserving communities, therefore the communities formed in $G_{AND}$ will be formed subsets of the communities in $G1$ and $G2$. Most importantly, due to the self preserving nature, no new grouping of vertices will be formed in $G_{AND}$. Therefore we can reconstruct the communities in $G_{AND}$ by simply taking the intersection of the communities of $G1$ and $G2$.

        \begin{figure}[h]
            \centering
            \includegraphics[width=0.6\textwidth]{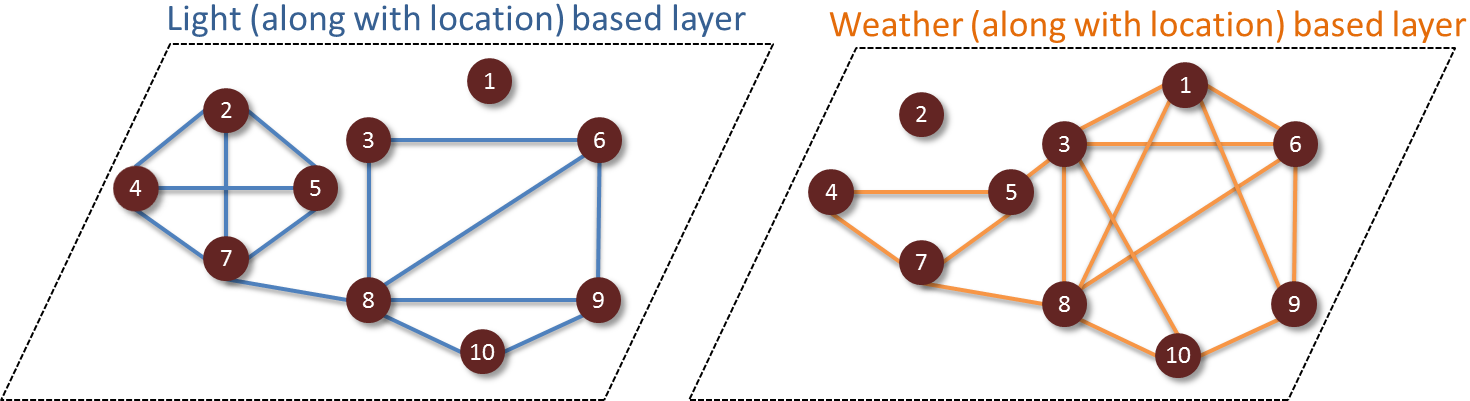}
            \caption{The layers (along with location) generated for a random set of accident instances}
            \label{fig:LightWeatherwithLocationLayers}
        \end{figure}

        \begin{figure}[h]
            \centering
            \includegraphics[width=0.6\textwidth]{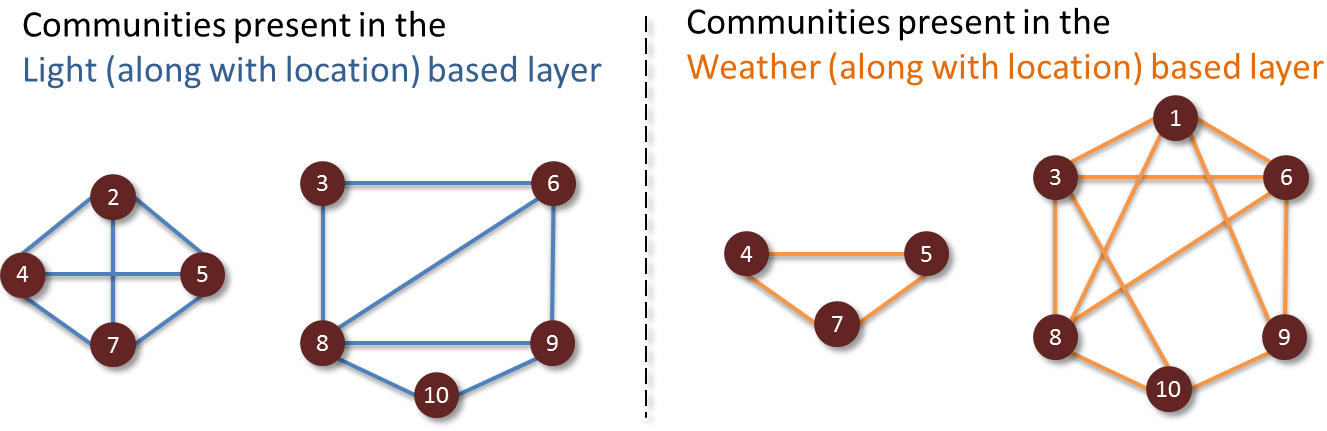}
            \caption{Actual Communities generated for the layers in Figure \ref{fig:LightWeatherwithLocationLayers}}
            \label{fig:Communities:LightWeather}
        \end{figure}

\begin{figure}
            \centering
            \includegraphics[width=0.6\textwidth]{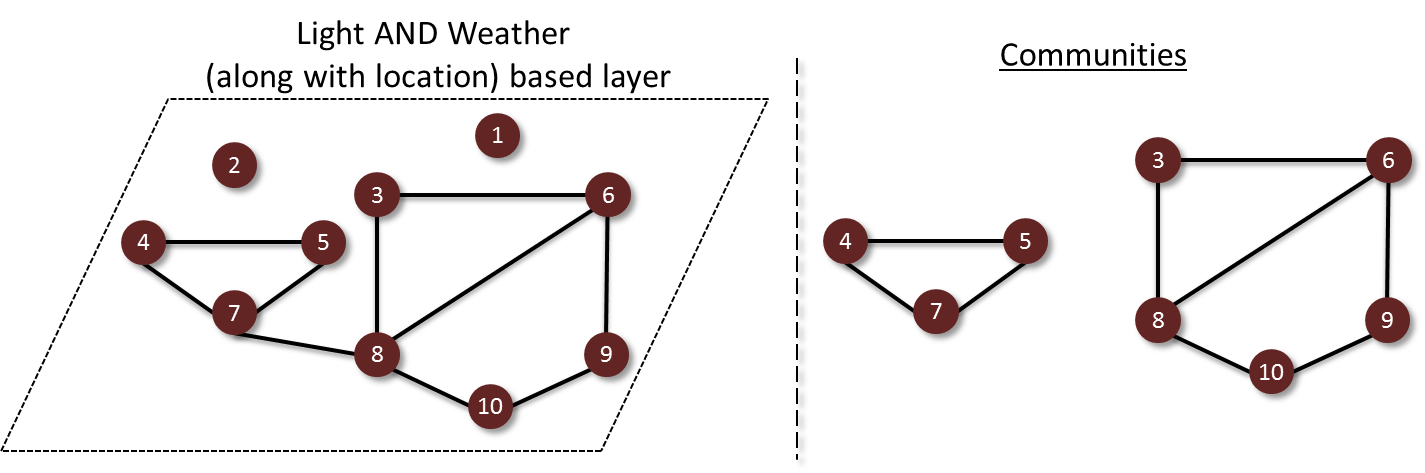}
            \caption{Actual Communities for the Light AND Weather (along with location) based layer}
            \label{fig:AllCommunities}
        \end{figure}

\begin{figure}
            \centering
            \includegraphics[width=0.6\textwidth]{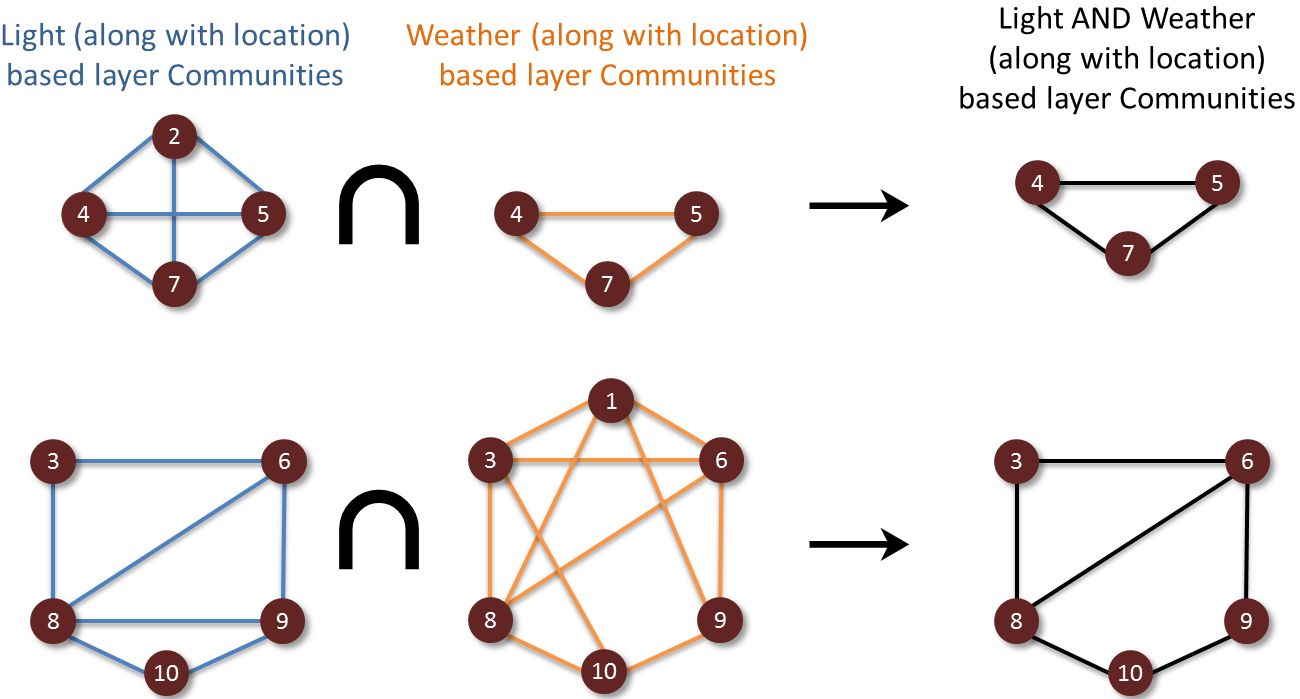}
            \caption{Pairwise intersection of the communities in the individual layers \emph{recreates} the communities of the AND composed layer}
            \label{fig:predictCommunities}
        \end{figure}

An example of such reconstruction is given in Figures \ref{fig:LightWeatherwithLocationLayers}- \ref{fig:predictCommunities}.
Figure \ref{fig:LightWeatherwithLocationLayers} shows two layers of networks and Figure \ref{fig:Communities:LightWeather} their corresponding communities, all of which are self-preserving. Figure \ref{fig:AllCommunities} shows the AND-composed network and the resultant communities. Figure \ref{fig:predictCommunities} shows that for this toy example, we can indeed reconstruct the communities for the AND-composed networks by taking the intersection of the  communities from the two separate networks.

\section{Experimental Results}
\label{sec:Experiments}
In this section we present our experimental results on composing networks with combined features and recreating the communities in these composed networks. Specifically, we i) construct user-defined individual layers, ii) perform Boolean compositions of the generated individual layers and iii) validate that the communities obtained by intersection of the individual layers are the same as the communities obtained by the composed layer.

\label{Experimental Setup}We use a dataset of road accidents that occurred in the United Kingdom in the year 2014 \cite{UKDataset2014}. Out of a total of 32 attributes captured for each accident, we use three nominal features - light conditions with domain as \emph{\{daylight, darkness: lights lit, darkness: lights unlit, darkness: no lighting, darkness: lighting unknown\}}, weather conditions with domain as \emph{\{fine + no high winds, raining + no high winds, snowing + no high winds, fine + high winds, raining + high winds, snowing + high winds, fog or mist, other\}} and road surface conditions with domain as \emph{\{dry, wet or damp, snow, frost or ice, flood, oil or diesel, mud\}} for the first three  individual layers ($L_1$, $L_2$ and $L_3$). The latitude and longitude values of accident location were grouped to form the geographical location based fourth layer, $L_4$ and time was the fifth layer $L_5$.

Our codes were implemented in C++ and were executed on a Linux based machine with 4 GB RAM, 500GB of local disk space and installed with UBUNTU 13.10. We used Infomap \cite{InfoMap2014} to detect communities in the networks, with a setting which assigns any node to at most one community.

\textbf{Generating the Layers per Feature:} Three layers of our network, light, weather and road conditions are of nominal type, therefore an edge is added if the values match exactly. For layers $L_4$ and $L_5$ the appropriate threshold has to be determined. There is a trade-off here, because too low a threshold can lead to loss of information, and too high a threshold leads to a dense network that is expensive to analyze.

\begin{figure}
    \centering
    \includegraphics[width=0.8\textwidth]{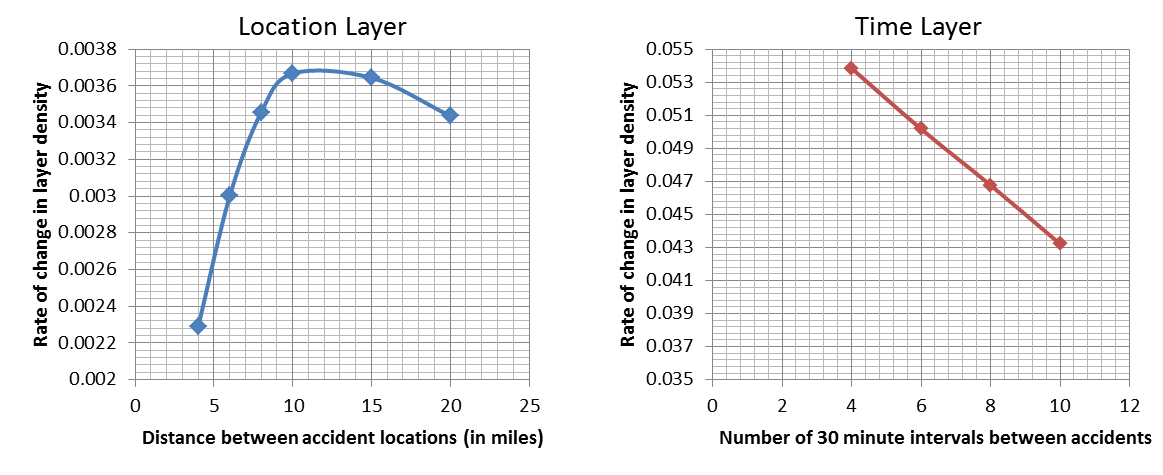}
    \caption{Variation in rate of change in layer density with threshold value}
    \label{fig:layerDensity}
\end{figure}

{\it Identifying appropriate thresholds:} To identify the appropriate threshold, we plotted different thresholds for distance ($L_4$) and time ($L_5$) layers versus the density of the layer at that threshold. Figure \ref{fig:layerDensity} shows that the change in layer density, peaks in the interval 10-12 miles for the distance layer and in the interval 3.5-4 for the time layer, which is divided into 30 minute slots. Based on this information we selected the threshold for the distance at 10 miles and threshold for time at 1.5 hours.

\begin{figure}[h]
    \centering
    \includegraphics[width=0.5\textwidth]{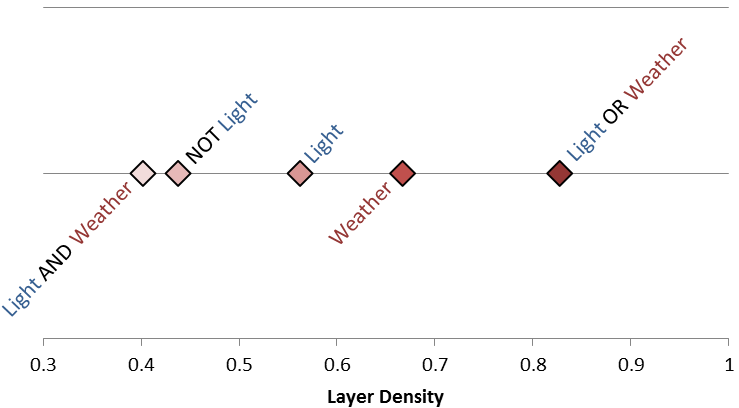}
    \caption{Distribution of densities for individual and composed layers for the accident event}
    \label{fig:compositionsLayer}
\end{figure}

\textbf{Density of basic and composed layers:} In Figure~\ref{fig:compositionsLayer}, we show the densities of the individual nominal layers (Light, Weather),  their different composed layers (AND and OR) and the complement graph for the Light layer (NOT) for a set of 1000 accidents from the dataset. The density of the AND-composed layer $(L_1$ \emph{AND} $L_2)$ will have an upper bound of the minimum density between $L_1$ and $L_2$, because it is formed of the intersection of the edges. Similarly, the union of all edges causes the density of the OR-composed layer $(L_1$ \emph{OR} $L_2)$ to have a lower bound of the maximum density between $L_1$ and $L_2$.

For our experiments we AND-composed each of the nominal layers with the distance and time layers to ensure that we are considering accidents within the same distance radius and same time interval. Thus when we refer to the Light layer we mean that it is Light AND Distance AND Time. We refer to the Weather and Road Condition layers similarly.

\begin{figure}
    \centering
    \includegraphics[width=\textwidth]{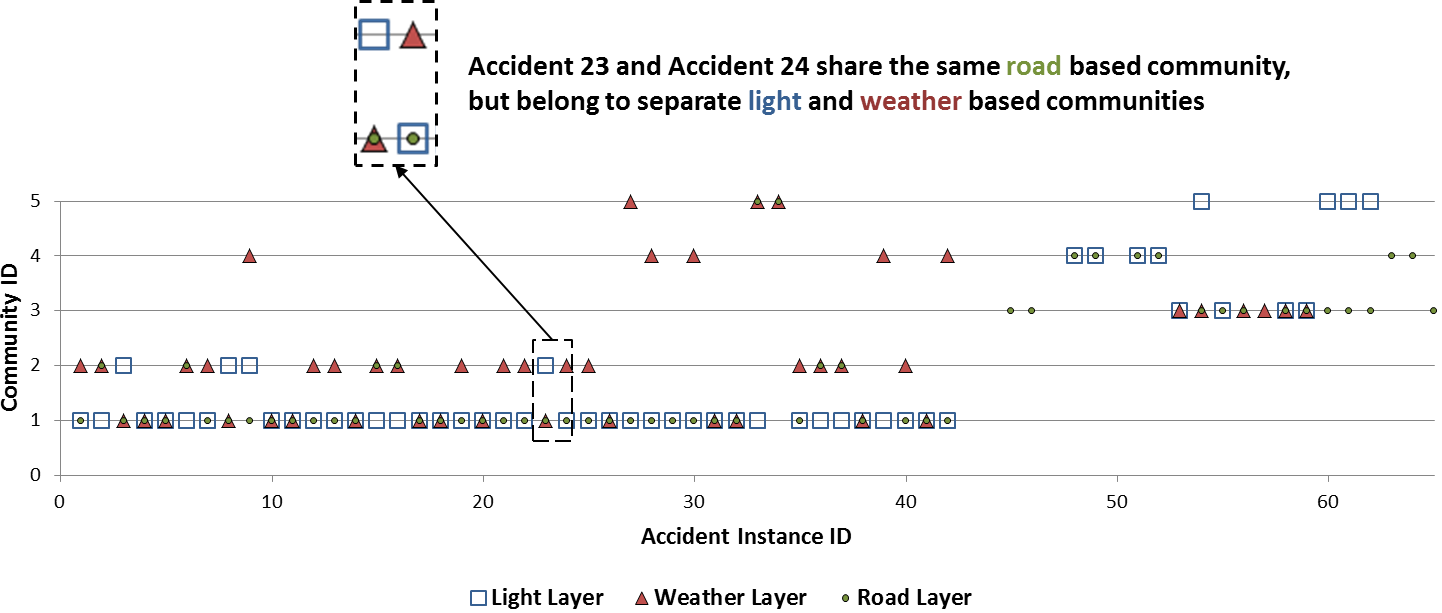}
    \caption{Minimal overlap among layer-wise communities for a snapshot of random accident instances}
    \label{fig:LightandWeatherComm}
\end{figure}

\textbf{Communities in the Individual and Composed Layers:} We now find the communities in the individual and composed layers to identify groups of accidents that are influenced by a similar set of features. In Figure \ref{fig:LightandWeatherComm} we plot a random set of a few accidents and their respective communities in the Light, Weather and Road layers. The X-axis shows the Id of the accident and the Y-axis the community to which the accident belongs. The squares, triangles and circles, indicate the communities obtained from the Light, Weather and Road layers respectively.

As can be seen from the figure, there are several accidents that are assigned to the same community by multiple layers. But there are certain accidents like accident number 23 and 24 that are assigned to the same community as per the Road layer, but to different communities as per the Light and Weather layers. The {\bf main takeaway is that there are accidents that are influenced by different subgroups of features.} 

\begin{figure}[h]
    \centering
    \includegraphics[width=\textwidth]{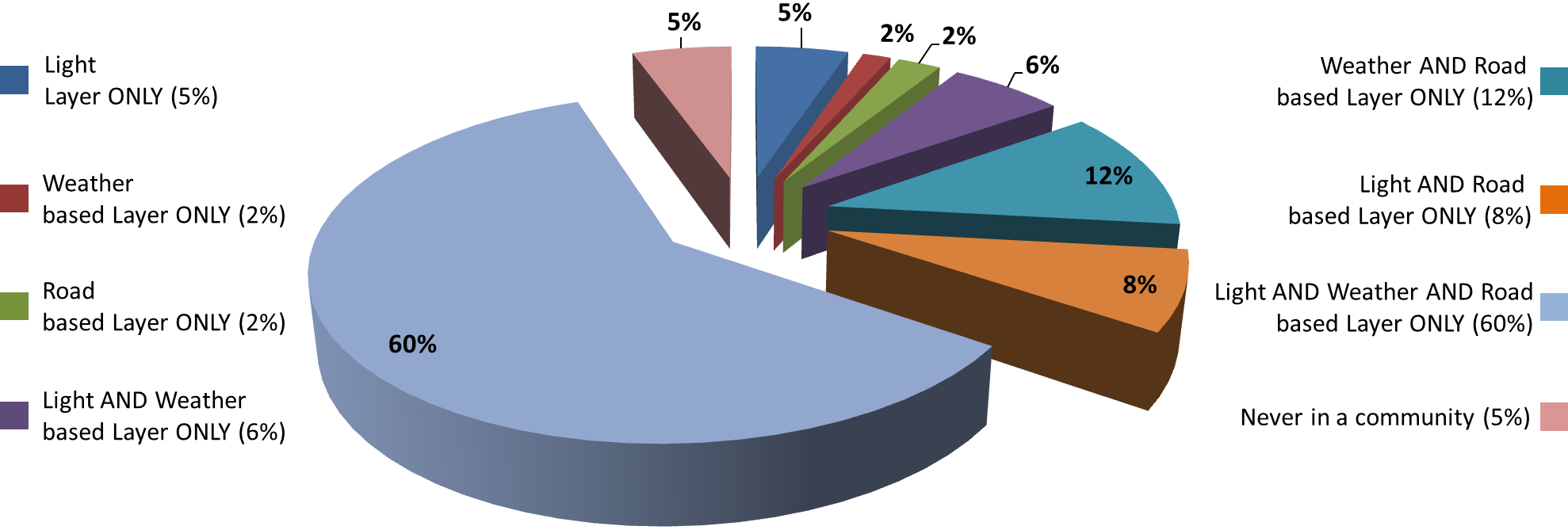}
    \caption{Percentage of instances that belong to some community with respect to individual or combination of features/layers (for 1000 random accident instances)}
    \label{fig:independentComm}
\end{figure}

We show the breakup of how 1000 accidents are grouped by the various individual and composed layers in Figure~\ref{fig:independentComm}. The pie-chart shows that $60\%$ of the accidents were grouped based on \emph{all the features}. $5\%$ of the accidents were not in any community. Therefore, a multilayer analysis of all features will lose information of the $35\%$ accidents that belonged to some community in other composed and individual layers. This highlights that {\bf it is equally important to analyze the individual layers and their various compositions.}

\textbf{Recreation of communities in the AND-composed networks:} As discussed earlier, computing communities from each of the composed networks is an expensive task. Here we show that we can successfully recreate the communities of the AND-composed layers, thus reducing the space to store the AND-composed layers and also the time.

We noticed that all the communities in the Light, Weather and Road  layers were self-preserving. Therefore we can recreate the communities in the AND-composed layers by simply intersecting the communities in the individual layers.

\begin{figure}[h]
    \centering
    \includegraphics[width=0.8\textwidth]{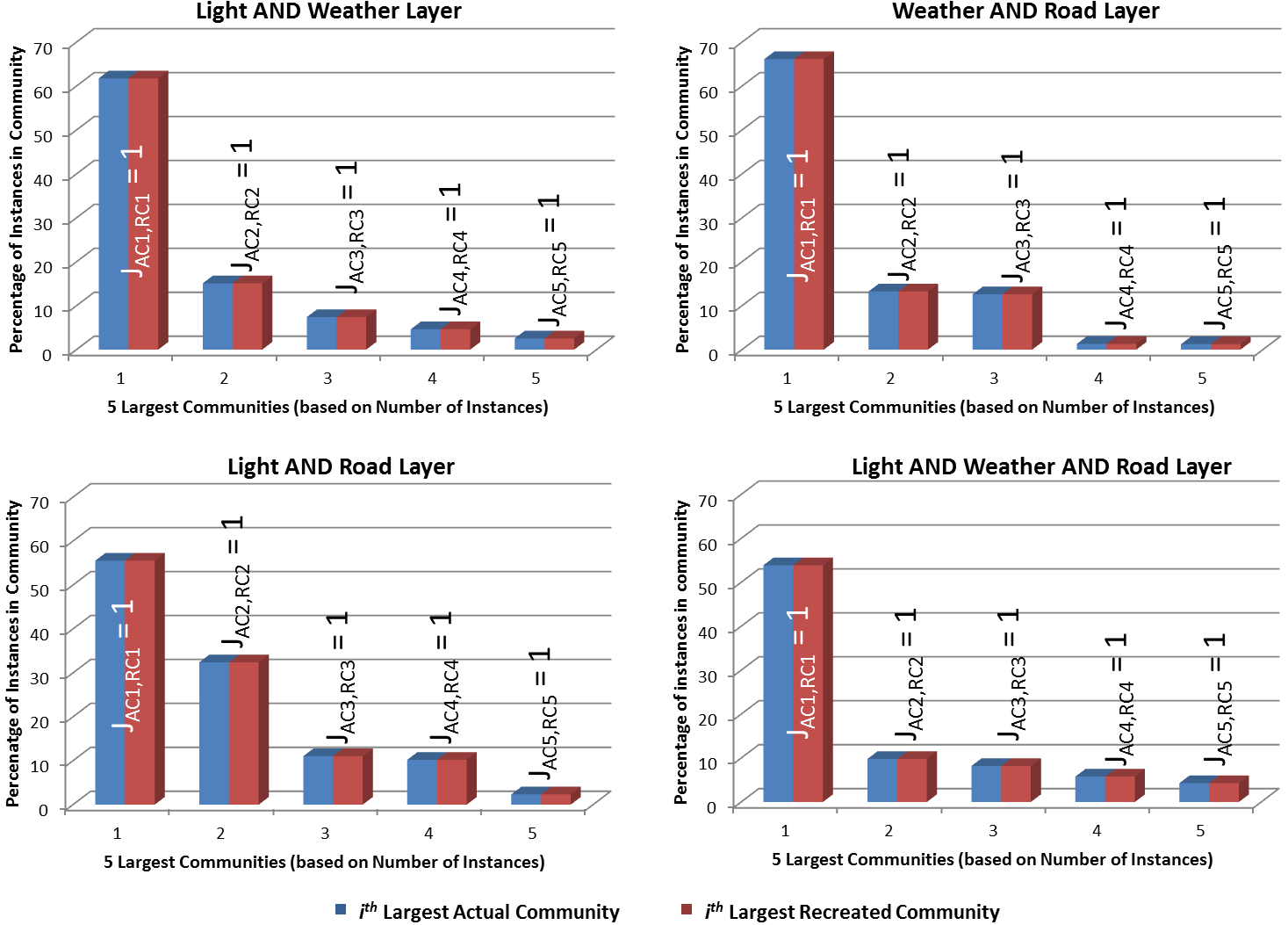}
    \caption{Comparison of the Jaccard Index ($J_{ACi, RCi}$) between the $i^{th}$ largest actual community and the $i^{th}$ largest recreated community, for various AND-compositions of Light, Weather and Road layers, for 3000 accidents.}
    \label{fig:InferCommunity1}
\end{figure}

Figure \ref{fig:InferCommunity1} shows the similarity between the communities created from the AND-composed networks (Light AND Weather, Light AND Road, Weather AND Road, and, Light AND Weather AND Road) and the communities recreated by intersecting the communities of the individual layers for 3000 accident instances. The similarity between the communities was computed using the Jaccard Index (J). For two sets $A$ and $B$, $J_{A, B}$ \emph{=} $\frac{A \cap B}{A \cup B}$. Thus a Jaccard value of 1 means that the two sets are identical. As can be seen from the sub-figures that the Jaccard value was 1 for the 5 largest communities for each of the AND-composed networks. We observed exactly the same results (J = 1) when testing on smaller datasets of 1000 and 2000 accident sets. This empirically validates that {\bf the communities in AND-composed networks can be successfully recreated by intersecting the communities in the individual networks.}

\begin{figure}
\centering
    \includegraphics[width=0.7\textwidth]{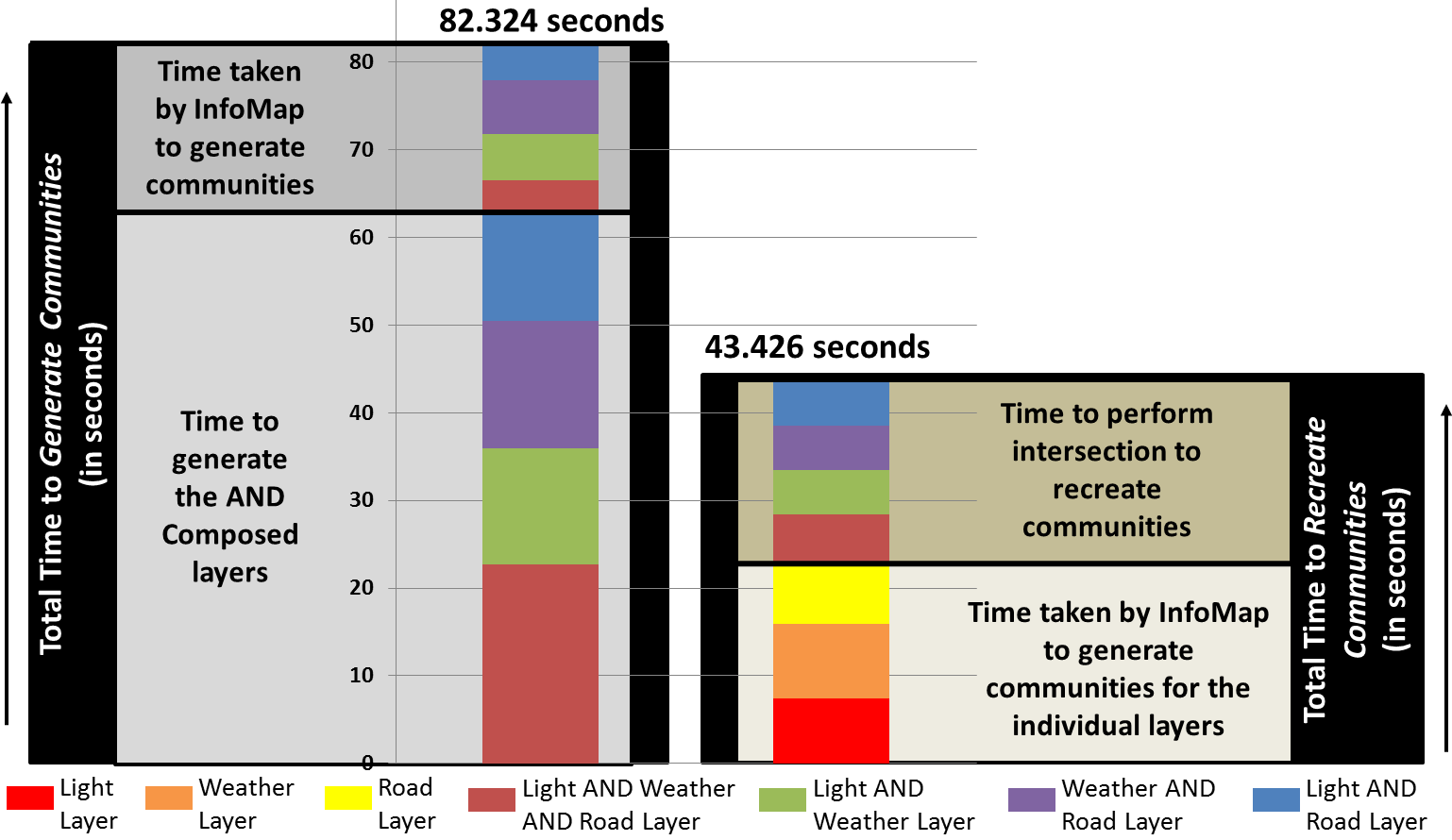}
   \caption{Comparison of time between generating and recreating the communities for AND-Composed Layers}
   \label{fig:CommTime}
\end{figure}

{\it Time to re-create the communities.} Figure \ref{fig:CommTime} compares the time to re-create the communities versus the time to generate them in the AND-composed networks on the 3000 accident dataset. To generate the communities in the individual layers it took {\bf 7.406 seconds}, {\bf 8.504 seconds} and {\bf 7.08 seconds}, for Light, Weather and Road Layers, respectively. After that it took {\bf 5.372 seconds}, {\bf 5.072 seconds}, {\bf 5.032 seconds} and {\bf 4.96 seconds} to perform the intersection of layer-wise communities to recreate the communities for Light AND Weather AND Road, Light AND Weather, Weather AND Road and Light AND Road composed layers, respectively.

In comparison it took {\bf 22.691 seconds}, {\bf 13.265 seconds}, {\bf 14.465 seconds} and {\bf 12.08 seconds} to create the above mentioned AND-composed layers and {\bf 3.992 seconds}, {\bf 5.271 seconds}, {\bf 6.122 seconds} and {\bf 4.438 seconds} to obtain the communities for them, respectively. Therefore, the recreation method was about 47\% faster, a total of {\bf 43.426 seconds} compared to a total of {\bf 82.324 seconds}. This is likely to improve further as the number of features increases.

These experimental results highlight that multilayered network is an effective tool for studying events associated with multiple data. They also show that in order to have a holistic understanding of the central event perspective-wise analysis is the key, that is we need to study the effect of all combinations of the features. Finally we show that recreating the communities from individual layers can reduce the computational costs of the analysis. 

\section{Conclusion and Future Extensions}
\label{sec:Conclusion}
This paper proposes a novel approach to model and analyze data fusion problems. This paper makes a case for multilayered analysis approach for multi-source, data fusion problems, its advantages, and composability aspects to improve modeling and computation aspects. Initial experimental results on real-world datasets have been very encouraging and empirically establish composability.

As future work, we plan on extending this work by introducing weighted and directed edges, modifying the composition schemes with respect to such type of edges, handling other types of features and distance metrics and come up with a generalized formulation for inferring communities for k-level composed layers/features based on single feature based communities, along with the theoretical analysis for this method's prediction accuracy.

\bibliographystyle{abbrv}
\bibliography{bibliography/santraResearch}

\end{document}